\documentclass[11pt,number,sort&compress]{elsarticle}

\usepackage[margin=1in]{geometry}

\usepackage{enumitem}
\setitemize{noitemsep,nolistsep,topsep=0pt,parsep=0pt,partopsep=0pt,leftmargin=*}

\usepackage[T1]{fontenc}

\usepackage[labelfont=bf, font=footnotesize, singlelinecheck=false, skip=10pt, tableposition=top, figureposition=top, labelsep=period]{caption}

\usepackage{titlesec}
\titlespacing\section{0pt}{12pt plus 4pt minus 2pt}{-\parskip}
\titlespacing\subsection{0pt}{0pt plus 4pt minus 2pt}{-\parskip}
\titleformat{\section}[block]{\normalfont\bfseries}{\thesection}{0ex}{}
\titleformat{\subsection}[block]{\normalfont\it}{\thesubsection}{0ex}{}

\renewcommand\theaffn{\arabic{affn}}
\makeatletter
\long\def\@@address[#1]#2{\g@addto@macro\elsaddress{%
    \def\baselinestretch{1}%
     \refstepcounter{affn}
     \xdef\@currentlabel{\theaffn}
     \elsLabel{#1}%
    \textsuperscript{\theaffn}#2\\}}

\renewcommand\@biblabel[1]{#1.} 

\def\ps@pprintTitle{%
     \let\@oddhead\@empty
     \let\@evenhead\@empty
     \let\@oddfoot\@empty
     \let\@evenfoot\@oddfoot}
\makeatother

\usepackage{pdfcomment}

\journal{Current Opinion in Biotechnology}
\begin{document}

\begin{frontmatter}
\title{Emerging whole-cell modeling principles and methods}
\author[inst,dept,equal]{Arthur P. Goldberg}
\author[inst,dept,equal]{Bal\'azs Szigeti}
\author[inst,dept,equal]{Yin Hoon Chew} 
\author[inst,dept,equal]{John A. P. Sekar}
\author[inst,dept]{Yosef D. Roth}
\author[inst,dept]{Jonathan R. Karr\corref{cor}} 

\address[inst]{Icahn Institute for Genomics and Multiscale Biology}
\address[dept]{Department of Genetics and Genomic Sciences\\Icahn School of Medicine at Mount Sinai, New York, New York, 10029, USA}
\address[equal]{These authors contributed equally to the work.}

\cortext[cor]{Corresponding author: Karr, Jonathan R (\href{mailto:karr@mssm.edu}{karr@mssm.edu})}

\begin{abstract}
Whole-cell computational models aim to predict cellular phenotypes from genotype by representing the entire genome, the structure and concentration of each molecular species, each molecular interaction, and the extracellular environment. Whole-cell models have great potential to transform bioscience, bioengineering, and medicine. However, numerous challenges remain to achieve whole-cell models. Nevertheless, researchers are beginning to leverage recent progress in measurement technology, bioinformatics, data sharing, rule-based modeling, and multi-algorithmic simulation to build the first whole-cell models. We anticipate that ongoing efforts to develop scalable whole-cell modeling tools will enable dramatically more comprehensive and more accurate models, including models of human cells.
\end{abstract}

\end{frontmatter}

\section*{Introduction}

Whole-cell (WC) computational models aim to predict cellular phenotypes from genotype and the environment by representing the function of each gene, gene product, and metabolite \cite{Karr2015CurrOpin}. WC models could unify our understanding of cell biology and enable researchers to perform \textit{in silico} experiments with complete control, scope, and resolution \cite{Tomita2001, Carrera2015}. WC models could also help bioengineers rationally design microorganisms that can produce useful chemicals and act as biosensors, and help physicians design personalized therapies tailored to each patient's genome.

Despite their potential, there is little consensus on how WC models should represent cells, what phenotypes WC models should predict, or how to achieve WC models. Nevertheless, we and others are beginning to leverage advances in measurement technology, bioinformatics, rule-based modeling, and multi-algorithmic simulation to develop WC models \cite{Tomita1999, Atlas2008, Roberts2011, Karr2012, Bordbar2015, Goldberg2016}. However, substantial work remains to achieve WC models \cite{szigeti2018blueprint, Macklin2014}.

To build consensus on WC modeling, we propose a set of key physical and chemical mechanisms that WC models should aim to represent, and a set of key phenotypes that WC models should aim to predict. We also summarize the experimental and computational progress that is making WC modeling feasible, and outline several technological advances that would help accelerate WC modeling.

Note, our proposals focus on defining WC models that are needed for research studies and applications such as bioengineering and personalized medicine which depend on understanding the molecular details of the majority of intracellular processes. However, research that depends on fewer intracellular processes could be served by smaller, more focused models.

\section*{Physics and chemistry that WC models should aim to represent}

\begin{figure}[tbh]
\centering
\includegraphics[width=\textwidth]{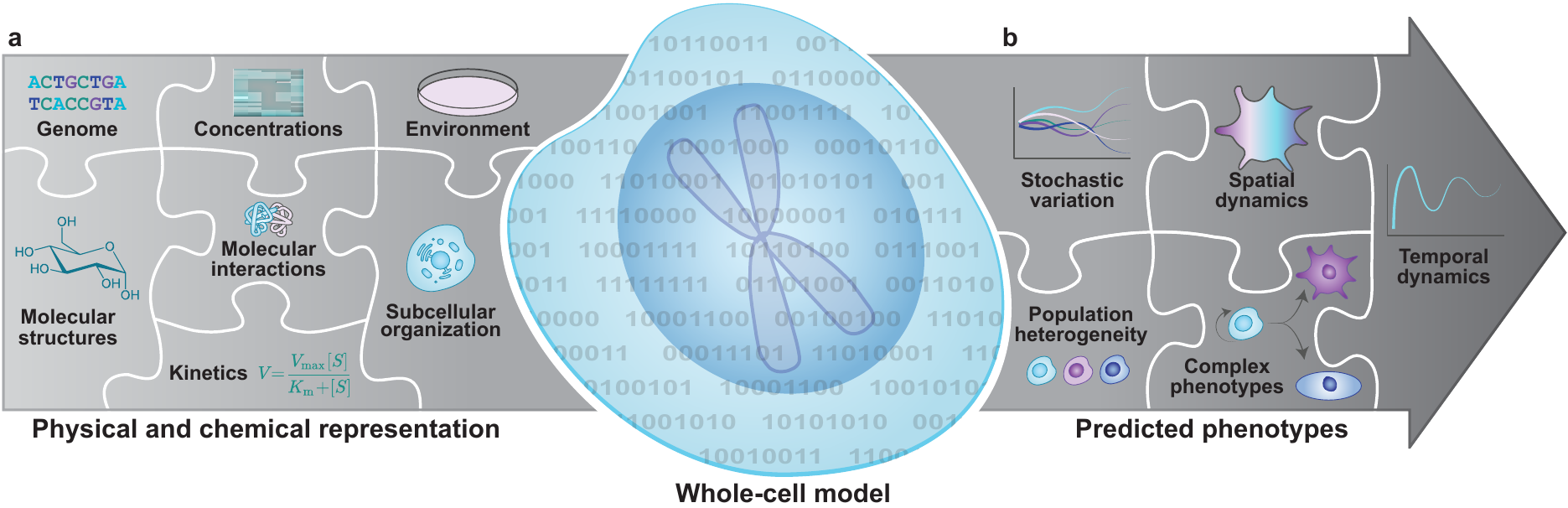}
\vspace{1em}
\caption{\label{fig:1} The physical and chemical mechanisms that WC models should aim to represent (a) and the phenotypes that WC models should aim to predict (b).}
\end{figure}

We propose that WC models aim to represent all of the chemical reactions in a cell and all of the physical processes that influence their rates (Figure~\ref{fig:1}a). This requires representing (a) the \textbf{sequence} of each chromosome, RNA, and protein; the location of each chromosomal feature, including each gene, operon, promoter, and terminator; and the location of each site on each RNA and protein; (b) the \textbf{structure} of each molecule, including atom-level information about small molecules, the domains and sites of macromolecules, and the subunit composition of complexes; (c) the \textbf{subcellular organization} of cells into organelles and microdomains; (d) the participants and effect of each \textbf{molecular interaction}, including the molecules that are consumed, produced, and transported, the molecular sites that are modified, and the bonds that are broken and formed, (e) the \textbf{kinetic parameters} of each interaction; (f) the \textbf{concentration} of each species in each organelle and microdomain; and (g) the concentration of each species in the \textbf{extracellular environment}. In addition, to enable WC models to be rigorously tested, each WC model should represent a single, well-defined experimental system. To minimize the complexity of WC models, we recommend modeling small, fast-growing, non-adherent, autonomous, self-renewing cells growing on defined, rich, homogeneous media. Together, this would enable WC models to describe how cellular behavior emerges from the combined function of each gene and genetic variant, and capture how cells respond to changes in their internal and external environments.

\section*{Phenotypes that WC models should aim to predict}

We also propose that WC models aim to predict the behavioral trajectories of single cells over their life cycles, with each simulation representing a different cell within a heterogeneous clonal population (Figure~\ref{fig:1}b). This should include behaviors within individual cells such as the \textbf{stochastic dynamics} of each molecular interaction; the \textbf{temporal dynamics} of the concentration of each species; the \textbf{spatial dynamics} of the concentration of each species in each organelle and microdomain; and \textbf{complex phenotypes} such as cell shape, growth rate, motility, and fate, as well as the \textbf{variation} in the behavior of single cells within clonal populations. Together, this would enable WC models to capture how stochastic and single-cell variation can generate phenotypic diversity; how a cell responds to external cues such as nutrients, growth factors and drugs; and how a cell coordinates critical events such as the G1/S transition. This would also enable WC models to generate predictions that could be embedded into higher-order multiscale models. For example, WC models could predict the timing and speed of chemotaxis, which could help multiscale models predict tumor metastasis.

\section*{Available resources}

Achieving WC models will require extensive data to constrain every parameter. Fortunately, measurement technology is rapidly advancing. Here, we review the latest methods for generating data for WC models, and highlight repositories and other resources that contain useful data for WC modeling.

\subsection*{Measurement methods}

Advances in single-cell and genomic measurement are rapidly generating data that could be used for WC modeling \cite{Macaulay2014, Altelaar2013, Fuhrer2015} (Table~S1). For example, Meth-Seq can assess epigenetic modifications \cite{Laird2010}, Hi-C can determine chromosome structures \cite{Dekker2013}, ChIP-seq can determine protein-DNA interactions \cite{Park2009}, fluorescence microscopy can determine protein localizations, mass-spectrometry can quantitate metabolite and protein concentrations, FISH \cite{Lee2014} and scRNA-seq \cite{Saliba2014} can quantitate the dynamics and single-cell variation of RNA abundances, and fluorescence microscopy and mass cytometry \cite{Bendall2012} can quantitate the dynamics and single-cell variation of protein abundances. In particular, WC models can be constrained by combining high-dimensional measurement methods with multiple genetic and environmental perturbations, frequent temporal observations, and cutting-edge distributed parameter estimation methods. However, substantial work remains to develop methods that can measure non-model organisms including small, slow-growing, and unculturable cells.

\subsection*{Data repositories}

Researchers are also rapidly aggregating much of the data needed for WC modeling into public repositories (Table~S2). For example, UniProt contains a multitude of information about proteins \cite{Uniprot2017}; BioCyc contain extensive information about interactions \cite{Caspi2016}; ECMDB \cite{Sajed2016}, ArrayExpress \cite{Kolesnikov2015}, and PaxDb \cite{Wang2015} contain metabolite, RNA, and protein abundances, respectively; and SABIO-RK contains kinetic parameters \cite{Wittig2012}. Furthermore, meta-databases such as \textit{Nucleic Acid Research}'s Database Summary contain lists of repositories \cite{Galperin2017}.

\subsection*{Prediction tools}

For certain types of data, accurate prediction tools can be superior to direct experimental evidence which may have incomplete coverage or may be limited to a small number of genotypes and environments. Currently, many tools can predict properties such as operons, RNA folds, and protein localizations (Table~S3). For example, PSORTb predicts the localization of bacterial proteins \cite{Yu2010}. However, many current prediction tools lack sufficient accuracy for WC modeling.

\subsection*{Published models}

WC models can also incorporate separately published models of individual pathways. Currently, there are several model repositories which contain numerous cell cycle, circadian rhythm, electrical signaling, signal transduction, and metabolism models (Table~S4--S5). However, most pathways such as RNA degradation do not yet have genome-scale dynamical models, many reported models are not publicly available, and it is difficult to merge most published models because they often use different assumptions and representations.

\section*{Emerging methods and tools}

\begin{figure}[tbh]
\centering
\includegraphics[width=\textwidth]{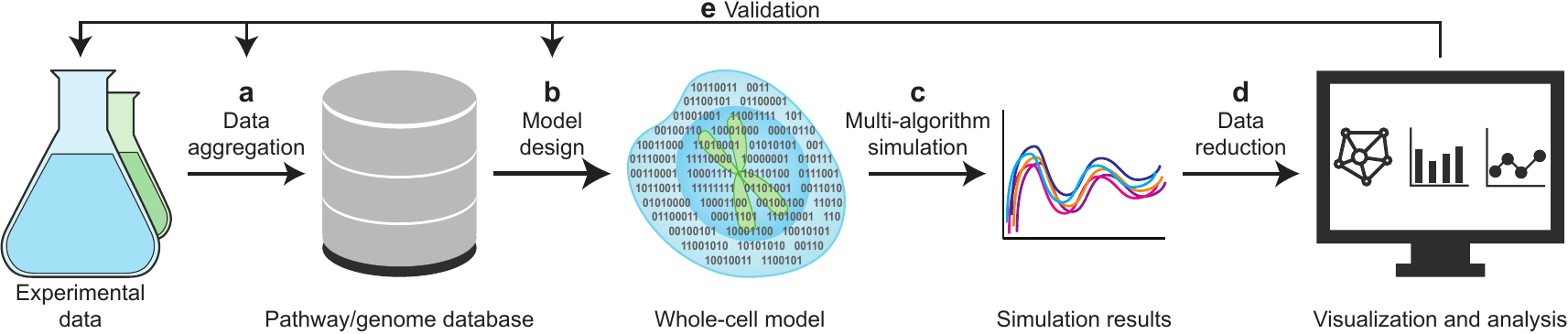}
\vspace{1em}
\caption{\label{fig:2} Emerging WC modeling methodology. (a) Data should be aggregated from thousands of publications, repositories, and prediction tools and organized into a PGDB. (b) Models should be designed, calibrated, and validated from PGDBs and described using rules. (c) Models should be simulated using parallel, network-free, multi-algorithmic simulators and their results should be stored in a database. (d) Simulation results should be visualized and analyzed. (e) Results should be validated by comparison to experimental measurements. Importantly, all of these steps should be collaborative.}
\end{figure}

Recent advances in data aggregation, model design, model representation, and simulation (Table~S6) are also rapidly making WC modeling feasible. We expect that ongoing efforts to adapt and combine these advances will accelerate WC modeling \cite{Goldberg2016} (Figure~\ref{fig:2}). Here, we summarize the most important emerging methods and tools for WC modeling.

\subsection*{Data aggregation and organization}

For optimal accuracy and scope, WC modeling should be tightly coupled with targeted experimentation. Nevertheless, we believe that WC modeling currently can be most cost-effectively advanced by leveraging the extensive array of public data. To make this public data usable for modeling, researchers are developing automated methods for extracting data from publications \cite{Cohen2015}, building central public repositories \cite{Pampel2013}, and creating tools for programmatically accessing repositories \cite{Cokelaer2013}. Pathway/genome database (PGDB) tools such as Pathway Tools \cite{Karp2016} are well-suited to organizing this data because they support structured representations of metabolites, DNA, RNA, proteins, and their interactions. However, they provide limited support for non-metabolic pathways and quantitative data. To overcome these limitations, we developed the WholeCellKB tool to organize data for WC modeling \cite{Karr2013}.

\subsection*{Scalable model design}

Several new tools can help researchers develop large models. For example, the Cell Collective facilitates collaborative model design \cite{Helikar2013}, MetaFlux facilitates the design of constraint-based models from PGDBs \cite{Latendresse2012}, PySB facilitates programmatic model construction \cite{Lopez2013}, SEEK facilitates model design from data tables \cite{Wolstencroft2015}, and Virtual Cell facilitates model design from KEGG and SABIO-RK \cite{Resasco2012}.

\subsection*{Model languages}

Researchers have developed several languages for representing biochemical models. SBML can represent several types of models including flux balance analysis models, deterministic dynamical models, and stochastic dynamical models \cite{Hucka2003}. Rule-based languages such as BioNetGen can efficiently describe the combinatorial complexity of protein-protein interactions \cite{Harris2016}.

\subsection*{Simulation}

Numerous tools can simulate biomodels. For example, COPASI \cite{Mendes2009} and Virtual Cell \cite{Resasco2012} support deterministic, stochastic, hybrid deterministic/stochastic, network-free, and spatial simulation; COBRApy supports constraint-based simulation \cite{Ebrahim2013}; and E-Cell supports multi-algorithmic simulation \cite{Dhar2006}.

\subsection*{Calibration}

New tools such as saCeSS \cite{Penas2017} support distributed calibration of large biochemical models. In addition, aerospace and mechanical engineers have developed methods for using reduced surrogate models to efficiently calibrate large models \cite{Forrester2009}. 

\subsection*{Verification}

Researchers have begun to adapt formal model checking techniques to biomodeling. For example, BioLab \cite{Clarke2008} and PRISM \cite{Kwiatkowska2011} can verify BioNetGen-encoded and SBML-encoded models, respectively.

\subsection*{Simulation results analysis}

Tools such as COPASI \cite{Mendes2009} and Virtual Cell \cite{Resasco2012} can visualize simulation results. We have developed the WholeCellSimDB \cite{Karr2014} simulation results database to help researchers organize, search, and share WC simulation results. We have also developed the WholeCellViz \cite{Lee2013} simulation results dashboard to help researchers visualize WC simulation results in their biological context.

\section*{Technological challenges}

Beyond these emerging tools, several technological advances are needed to enable WC models. Here, we summarize the most critically needed technologies.

\subsection*{Experimental measurement}

While substantial data about cellular populations already exists, additional data would enable better models. In particular, we need metabolome-wide and proteome-wide measurement technologies that can quantitate the dynamics and single-cell variation of each metabolite and protein. Additionally, we need technologies that can measure kinetic parameters at the interactome scale and technologies that can measure cellular phenotypes across multiple genetic and environmental conditions. Furthermore, to enable WC models of a broad range of organisms, we also need technologies that can measure non-model organisms, including small, slow-growing, motile, and unculturable organisms.

\subsection*{Prediction tools}

While existing tools can predict many properties of metabolites, DNA, RNA, and proteins, additional tools are needed to accurately predict the molecular effects of insertions, deletions, and structural variants. Such tools would help WC models design microbial genomes and predict the phenotypes of individual patients.

\subsection*{Data aggregation}

As described above, extensive data is now available for WC modeling. However, this data is scattered across many repositories and publications; spans a wide range of data types, organisms, and environments; is described using inconsistent identifiers and units; and often is not annotated or normalized. To make this data more usable for modeling, we are developing a framework for aggregating data from repositories; merging data from multiple species, environmental conditions, and experimental procedures; standardizing data to common units; and identifying the most relevant data for a model.

\subsection*{Scalable, data-driven model design}

To scale WC modeling, we need tools for collaboratively building large models directly from experimental data, recording how data is used to build models, and identifying gaps and inconsistencies in models. As described above, several tools support each of these functions. To accelerate WC modeling, the field must develop an extensible platform that supports all of these functions at the scale required for WC modeling.

\subsection*{Rule-based model representation}

Several languages can represent individual biological processes, but no existing language supports all of the biological processes that WC models must represent \cite{Waltemath2016, Medley2016}. To overcome this limitation, we are developing a rule-based language that can represent each molecular species at multiple levels of granularity (for example, as a single species, as a set of sites, and as a sequence); the combinatorial complexity of each molecular species and interaction; composite, multi-algorithmic models; and the data used to build models.

\subsection*{Scalable multi-algorithmic simulation}

Simulating WC models requires a simulator that supports both network-free interpretation of rule-based model descriptions and multi-algorithmic co-simulation of submodels that are described using different simulation algorithms. However, no existing simulator supports both network-free and multi-algorithmic simulation. To scalably simulate WC models, we are using Rete algorithms and parallel discrete event simulation to develop a parallel, network-free, multi-algorithmic simulator \cite{Goldberg2016}.

\subsection*{Calibration and verification}

Scalable tools are needed to calibrate and verify WC models. Although we and others have begun to explore surrogate strategies for efficiently calibrating and validating WC models \cite{Karr2015PLoS}, further work is needed to formalize these methods.

\subsection*{Simulation analysis}

We and others have developed tools for organizing and visualizing simulation results, but they provided limited support for large datasets or custom visualizations such as pathway maps. To visualize WC simulation results, researchers should use distributed database and data processing technologies to search and reduce simulation results, standard visualization grammars to enable flexible and custom visualizations, and high-performance visualization toolkits to handle terabyte-scale simulation results.

\subsection*{Collaboration}

Ultimately, achieving WC models will require extensive teamwork. To facilitate collaboration, the field must develop collaborative model design tools, version control systems for models, standards for annotating and verifying submodels, and protocols for merging separately developed submodels.

\section*{Conclusion}

WC models have great potential to advance bioscience, bioengineering, and medicine. However, significant challenges remain to achieve WC models. To advance WC modeling, we have proposed how WC models should represent cells and the phenotypes that WC models should predict, and summarized the best emerging methods and resources. We have also outlined several technological solutions to the most immediate WC modeling challenges. Specifically, we must develop new tools for scalably and collaboratively designing, simulating, calibrating, validating and analyzing models. We must also develop new methods for measuring the dynamics and single-cell variation of the metabolome and proteome and for measuring kinetic parameters at the interactome scale. Despite these challenges, we and others are building the first WC models, developing the first WC modeling tools, and beginning to form a WC modeling community \cite{Karr2015PLoS, Waltemath2016}. We anticipate that these efforts will enable comprehensive models of cells.

\section*{Acknowledgements}

We thank Saahith Pochiraju for critical feedback. This work was supported by a National Institute of Health MIRA award [grant number 1 R35 GM119771-01]; a National Science Foundation INSPIRE award [grant number 1649014]; and the National Science Foundation / ERASynBio [grant numbers 1548123, 335672].

\section*{References and recommended reading}
Papers of particular interest, published within the period of review, have been highlighted as:
\begin{description}[nosep,labelindent=1em]
\item[$\bullet$]~~~of special interest
\item[$\bullet\bullet$]~of outstanding interest
\end{description}

\bibliographystyle{elsarticle-num}
\bibliography{References}

\end{document}